\documentstyle[aps,prd]{revtex}
\begin{document}
\draft
\title{Effect of gravitational radiation reaction on\\
nonequatorial orbits around a Kerr black hole}
\author{Fintan D. Ryan}
\address{Theoretical Astrophysics, California Institute of Technology,
	Pasadena, California 91125}
\date{November 22, 1995}
\maketitle
\begin{abstract}

The effect of  gravitational  radiation reaction  on orbits  around  a
spinning black hole is analyzed.  Such  orbits possess three constants
of motion: $\iota$, $e$,  and $a$, which  correspond, in the Newtonian
limit of the orbit  being an ellipse, to  the inclination angle of the
orbital plane to the hole's  equatorial  plane, the eccentricity,  and
the semi-major axis  length, respectively.  First,  it is  argued that
circular orbits  ($e=0$) remain circular under gravitational radiation
reaction.  Second, for elliptical  orbits (removing the restriction of
$e=0$), the evolution of $\iota$, $e$, and $a$  is computed to leading
order in $S$ (the magnitude of the spin angular  momentum of the hole)
and  in $M/a$,  where  $M$ is  the mass   of the black   hole.  As $a$
decreases, $\iota$ increases and $e$ decreases.

\end{abstract}
\pacs{PACS numbers: 04.25.Nx, 04.30.Db}

\narrowtext

\section{Introduction}
\label{sec:intro}

The   Earth-based   Laser      Interferometer   Gravitational     Wave
Observatory--(LIGO-)VIRGO~\cite{sci1,sci2}  network  of  gravitational
wave detectors  (which  is now  under construction)  and  the European
Space Agency's planned space-based  Laser Interferometer Space Antenna
(LISA)~\cite{lisa}   will be used   to   search  for  and study    the
gravitational waves  from  ``particles'',  such as neutron   stars and
small black holes, spiraling into massive black holes  (mass $M$ up to
$\sim 300 M_\odot$  for LIGO/VIRGO and  up to $\sim 10^7  M_\odot$ for
LISA).  To search for the  inspiral waves and extract the  information
they carry will require templates based on theoretical calculations of
the emitted waveforms, which in turn  require a detailed understanding
of how radiation reaction influences the orbital evolution.

When the orbital plane of  the particle is  inclined to the equatorial
plane of   a  spinning hole, only  one   method has been  successfully
implemented to deduce how radiation  reaction influences the evolution
of the   orbit's ``Carter constant''~\cite{carter,mtw},  which governs
the orbital  shape and inclination  angle.  This method, which  uses a
``post-Newtonian''   gravitational   radiation  reaction  force,   was
described  in a previous  paper~\cite{circ}, but there only applied to
``circular   orbits''  (orbits of    constant Boyer-Lindquist   radial
coordinate $r$)  for simplicity.  This follow-up  paper has a two-fold
purpose:  First,  in Sec.~\ref{sec:orb}, we   will argue that circular
orbits   remain    circular under  gravitational   radiation reaction.
Second,  in Sec.~\ref{sec:reac},   we  will compute the   evolution of
elliptical orbits under radiation  reaction, but only to leading order
in $S$, the magnitude of the spin  angular momentum of the black hole,
and leading order in $M/a$, where $M$ is the black hole's mass and $a$
is the size of the orbit, as  defined more precisely below.  (Here and
throughout, units with $G=c=1$ are used.)

\section{Evolution of circular orbits}
\label{sec:orb}

Several  years   ago, Ori~\cite{ori} put   forth  the  conjecture that
circular   orbits in  the  Kerr  metric  remain   circular even  under
gravitational radiation reaction.  Here, we will argue in favor of the
conjecture.  We will start by reviewing  some properties of elliptical
and circular orbits  in the Kerr metric.    Then we will argue  that a
circular orbit  and  the reaction  force acting  on it have  a type of
reflection symmetry that ensures that the orbit remains circular under
radiation reaction,  in the limit  of the particle's mass  being small
compared to the hole's mass.

In the absence of  gravitational radiation, the  geodesic motion  of a
particle   in   orbit around a  Kerr  black    hole is  well-known and
discussed, for example, in Sec.~33.6 of Ref.~\cite{mtw}.  The location
of the  particle can be described  in Boyer-Lindquist coordinates $r$,
$\theta$,  $\phi$,  and $t$.   The  orbit  can  be described  by three
constants  of motion: the energy  $E$,  the angular momentum along the
hole's spin axis  $L_z$, and the Carter  constant $Q$.  The particle's
rest mass $\mu$ can be counted as another constant of the motion.  The
energy $E$ is defined as the relativistic energy of the particle minus
its rest mass, so that ``$E-\mu$'' in  the language of Ref.~\cite{mtw}
corresponds to ``$E$'' here.  We  will restrict to bound orbits,  that
is $E<0$ and, as a consequence (see Ref.~\cite{carter}), $Q \geq 0$.

An  interesting feature  of   the   Kerr metric  in    Boyer-Lindquist
coordinates is  the  existence  of nonequatorial, circular,   geodesic
orbits.  These orbits  are  circular in  the  sense that the  particle
maintains a   constant Boyer-Lindquist  coordinate  $r$; however,  the
plane of the circular  orbit is not fixed  but rather precesses around
the hole's spin axis.  Such orbits exist  and are stable for values of
$E$, $L_z$, and $Q$ that give $R=0$,  $\partial R/ \partial r =0$, and
$\partial^2  R/ \partial  r^2 <  0$,  where  $R$ {\bf (}see Eq.~(33.33c)  of
Ref.~\cite{mtw}{\bf )} is defined by
\begin{eqnarray}
\label{eq:rfunc}
R=&&\left[(E+\mu)(r^2 + S^2/M^2)-L_z S/M \right]^2\nonumber\\
&&- (r^2 - 2M r +S^2/M^2)\nonumber\\
&&\times \left[\mu^2 r^2 +(L_z - S \mu/M - S E/M)^2 + Q \right].
\end{eqnarray}

For an arbitrary orbit with constants $E$, $L_z$, and $Q$, there might
be some other  energy $\bar E  \leq E$ ($\bar  E$ depends on $L_z$ and
$Q$) such that, if the orbit had energy  $\bar E (L_z, Q)$ rather than
$E$, the orbit would  be circular and  stable.  In such  a case, as an
alternative  set of  constants to  $E$, $L_z$,  and $Q$, the constants
$\iota$, $e$, and $a$ can be defined as follows:

\begin{mathletters}
\label{eq:defofiea}
\begin{eqnarray}
\label{eq:defofi}
\cos \iota \equiv&& \frac{L_z}{\left(Q+L_z^2\right)^{1/2}},\\
1-e^2 \equiv&& \frac{E}{\bar E},\\
a \equiv&&\frac{\bar E \bar r}{E}.
\end{eqnarray}
Here $\bar r = \bar  r (L_z, Q)$ is  the radius of the circular  orbit
with constants $\bar E$, $L_z$, and $Q$.  Note that $a$ should not be
confused  with  the conventional notation  for the  spin  of the black
hole, which is $S$ here.
\end{mathletters}

The positive root in Eq.~(\ref{eq:defofi}) or in any other square root
is always chosen.  We choose the angle $\iota$ to lie in the range $0
\leq \iota \leq \pi$, so that $\iota <  \pi/2$ corresponds to an orbit
co-rotating relative to     the     spin and $\iota    >    \pi/2$  to
counter-rotating.  Also, $e$ is chosen as nonnegative.

This   set of  constants $\iota$,  $e$,  and  $a$  has the  conceptual
advantage that in the Newtonian  limit of large $a$,  the orbit of the
particle is  an ellipse of eccentricity  $e$ and semimajor axis length
$a$,  on a   plane   with inclination   angle $\iota$  to   the hole's
equatorial  plane.  When not   in  the Newtonian  limit,  interpreting
$\iota$,  $e$, and  $a$ as  the  inclination angle, eccentricity,  and
semimajor  axis length  must be done  with  the caveat  that since the
orbit is  not  an ellipse,  then  words such  as ``eccentricity''  are
subject to a modified interpretation and can be misleading.

Even   though the particle's  motion is  complicated   when not in the
Newtonian limit,  some of the parameters  that describe the particle's
motion need not be specified.  For example,  we are not concerned with
the value of $\phi$ or $t$, because making a $\phi$ or $t$ translation
does not  change the physics  in   the axisymmetric,  stationary  Kerr
metric.  Another symmetry is  that  if the  orbital motion  is flipped
over   the hole's equatorial plane,   i.e., $\theta(t)$ is replaced by
$\pi-\theta(t)$, the motion can  be  considered  the same.   All  such
$\phi$ and $t$  translations and $\theta$  reflections leave the shape
of the orbit unchanged.

We can think of the particle as undergoing oscillatory, coupled motion
in  the  $r$ and   $\theta$ directions.   We  define  one {\it orbital
revolution} to  be one oscillation cycle as  measured  by the $\theta$
motion.  Given any chosen starting point of an orbital revolution with
coordinate  $\theta_0$, the revolution  can   be broken into  two half
revolutions, the first when the particle goes  from $\theta_0$ to $\pi
- \theta_0$ half  a  $\theta$-cycle later,  and  the  second when  the
particle goes from $\pi  - \theta_0$ back  to $\theta_0$  another half
$\theta$-cycle later.  [Because of the coupling of the $r$ motion with
the  $\theta$ motion,  the $\theta$ motion  does not  peak at the same
extrema  every cycle.  Therefore,  $|\pi/2-\theta_0|$ has to be chosen
small enough that the orbit does indeed  go through $\pi-\theta_0$ and
$\theta_0$ in  the  following cycle.   However, this is  a  very minor
restriction for the rest of Sec.~\ref{sec:orb},  where in proving that
circular  orbits stay circular, we  only consider  circular and almost
circular orbits (we do not have to consider generally eccentric orbits
since we know    that  a circular   orbit cannot  immediately   become
generally eccentric without first  being slightly eccentric).  In such
case,  the  peaks of the  $\theta$ motion  are   almost the same every
cycle.]

Now we  consider the effect of  gravitational radiation reaction on an
orbit.   We assume that the  rest mass $\mu$ is  small  enough for the
adiabatic  approximation to   hold:   the timescale  of  gravitational
radiation  reaction is  much longer than  any  other timescale  in the
problem.  Then the   particle moves very  nearly  on a   geodesic path
characterized by the constants of   motion $\iota$, $e$, and $a$;  and
only on a very long timescale (which varies like $1/\mu$ as $\mu
\rightarrow 0$, because   the radiation reaction  acceleration  scales
like $\mu$) is  this  motion substantially modified  by  gravitational
radiation reaction.

We now  consider, for an  orbit slowly  inspiraling  due to  radiation
reaction,   an   orbital revolution   that   satisfies  the  following
condition, to which   we   give the name {\it    reflection symmetry}:
Consider  the point  on  the orbit  that is  at  the  beginning of the
orbital revolution.  Denote by $r_0$, $\theta_0$, $\dot r_0$, $\dot
\theta_0$, and  $\dot \phi_0$ the  Boyer-Lindquist spatial coordinates
of that  point and their  time derivatives.   (Here and throughout, an
overdot represents $d/dt$.)  Then there are  two other locations later
on the path with coordinates
\begin{mathletters}
\label{eq:crit}
\begin{eqnarray}
r_n      = &&r_0+ \mu n \tilde r + {\rm h.o.},\\
\theta_n -\pi/2= &&(-1)^{2n}(\theta_0-\pi/2+\mu n \tilde \theta) +{\rm h.o.},\\
\dot r_n = &&\dot r_0 + \mu n \dot{\tilde r} + {\rm h.o.},\\
\dot \theta_n=&&\dot \theta_0 + \mu n \dot{\tilde \theta} + {\rm h.o.},\\
\dot \phi_n=&&\dot \phi_0 + \mu n \dot{\tilde \phi} + {\rm h.o.},
\end{eqnarray}
for $n={\case 1/2}$ (a half revolution after $n=0$)  and $n=1$ (a full
orbital revolution after  $n=0$).  The functions  with  tildes are not
functions of $\mu$.  The ``h.o.''~terms are  any terms that go to zero
faster than $\mu$ as $\mu \rightarrow 0$  ({\it higher order} in $\mu$
than linear).
\end{mathletters}

Because of the  initial conditions at  the beginning of the first  and
second half revolutions  (at $n=0$ and $n={\case 1/2}$, respectively),
the shape of the first half revolution  (the path connecting the $n=0$
and $n={\case 1/2}$  locations) deviates from  the shape of the second
half  revolution (the  path connecting  the $n={\case  1/2}$ and $n=1$
locations) by a path  deviation of order $\mu$.   Of course, these two
paths also differ  by a $\phi$ translation, a  $t$ translation,  and a
reflection across  the equatorial plane.   But as we  discussed above,
these   are unimportant differences because   of  the symmetries;  the
shapes of the paths are the same.

Now that we have written Eqs.~(\ref{eq:crit}), we temporarily (for the
remainder of  this paragraph)  go  back to  the case of   no radiation
reaction, i.e., we set to  zero the $\mu$  terms and the h.o.~terms in
Eqs.~(\ref{eq:crit}).   Clearly,   a   circular   orbital   revolution
satisfies Eqs.~(\ref{eq:crit}) for  any initial  $n=0$ location chosen
on the circular  orbit.  But could there  be an eccentric  orbit which
also satisfies Eqs.~(\ref{eq:crit})?    The answer is negative, as  we
shall now show.  A slightly eccentric orbit (one with the value of $e$
small enough  that $e^2$  terms are negligible)  would have  the same
$\theta (t)$ and $\phi(t)$ motion regardless  of the value of $e$, but
$r-\bar r$  would  oscillate with an  amplitude   proportional to $e$.
This  can be   verified   from  the Kerr-metric  geodesic   equations,
Eqs.~(33.32)  of Ref.~\cite{mtw}.     In  the  Newtonian  limit,   the
oscillation of $r-\bar r$ would be periodic with  the same period that
$\theta(t)$ has, but when not in the Newtonian  limit the $\theta$ and
$r$ oscillations would have different periods.  If an orbit were to be
reflection  symmetric, then  $r-\bar r$  would have  to have  the same
value when the orbital motion is at  $\theta_0$ as it  does when it is
at $\pi-\theta_0$  at the next value of  $n$.  This would require that
either $r-\bar  r$ oscillate at a  frequency  that is  an even integer
multiple of  the  $\theta$ oscillation  frequency, or $r-\bar  r$ have
zero amplitude  (a circular orbit).  The former  is never the case, as
can be verified  by  numerically~\cite{mathematica} examining circular
orbits in   the  Kerr metric  over  the  space of  possible physically
acceptable  values of $S$,  $L_z$ and $Q$.  The   fact that $r-\bar r$
does  not resonate with  an  even multiple of  the $\theta$  frequency
implies that a     slightly  eccentric orbit   cannot   be  reflection
symmetric.

Now we shall return  to the case of  interest: that with gravitational
radiation reaction.  What  precisely do we   refer to when  we discuss
circular orbital   revolutions, when the   orbital revolution  is  not
actually circular but is slowly inspiraling?   A good, but not unique,
definition is  one that  agrees with  the  result  in  the case  of no
radiation  reaction: We define  that a circular  orbital revolution is
one that satisfies  Eqs.~(\ref{eq:crit}),  while an eccentric  orbital
revolution is one that does not (at  least for slight eccentricity: as
mentioned  above, we are  not considering generally eccentric orbits).
An orbital  revolution   with weak  radiation reaction is   defined as
circular if and only if it is reflection symmetric.

We now  consider starting with an initial  orbital  revolution that is
circular, or equivalently,   that is reflection symmetric, i.e.,  that
satisfies  conditions~(\ref{eq:crit}).  For  small $\mu$, ignoring the
h.o.~(higher than $\mu$)  corrections, we would  expect that the third
half revolution (the first half  of the next orbital revolution) would
have a shape that deviates from that of the  second half revolution by
the  same amount as  the shape of  the second half revolution deviates
from   that      of  the  first.   We      expect  this,  because from
conditions~(\ref{eq:crit}) above, the  initial conditions of the third
half revolution differ from those of the second by the same amount (to
linear order in $\mu$) as those of the second differ from those of the
first;  and the   acceleration on the  particle    should similarly be
equally  (also to     linear   order in  $\mu$)     different  between
corresponding locations  on the second and  third half  revolutions as
between corresponding  locations on the first  and second.   The orbit
remains circular for the additional half  revolution.  If there is any
eccentricity    added,  it is in  the    h.o.~terms, but  in  the $\mu
\rightarrow 0$  limit, this is ignorable  compared to the shrinking of
the orbit, which varies like $\mu$ (the terms involving tildes).

We can repeat the above argument to  get the shape  of the fourth half
revolution, as  well as the fifth, sixth,  etc.  In fact, the argument
can  be  repeated to any   chosen  number, $n_{\rm  max}$, of  orbital
revolutions, as long as that chosen number  does not go to infinity as
$\mu  \rightarrow 0$; for if  it did, then we  would not be guaranteed
that  after  the infinite    number  $n_{\rm  max}$  of  orbits,   the
h.o.~corrections   of the above  paragraph  would  be ignorable.   For
example, we  could choose $n_{\rm max}$ to  be $100$, but we could not
choose it to be  $100 M/\mu$.  The  orbit remains reflection symmetric
(or equivalently, it  remains circular) for  $n$ up to $n_{\rm  max}$,
where $n$ increments  by ${\case 1/2}$.  In   other words, there  is a
location,  with  coordinates   $r_n$, $\theta_n$   $\dot  r_n$,  $\dot
\theta_n$, and $\dot  \phi_n$, satisfying Eqs.~(\ref{eq:crit}) for any
$n$ up to $n_{\rm max}$.

The constants of motion $E$, $L_z$, and $Q$ (or equivalently, $\iota$,
$e$, and   $a$) evolve in such   a  way that in    going from $n=0$ to
$n=n_{\rm  max}$ a circular orbit remains  circular.  By assigning new
values of $r_0$, $\theta_0$, $\dot  r_0$,  $\dot \theta_0$, and  $\dot
\phi_0$ as  the old  $r_{n_{\rm max}}$, $\theta_{n_{\rm  max}}$, $\dot
r_{n_{\rm max}}$, $\dot \theta_{n_{\rm  max}}$, and $\dot \phi_{n_{\rm
max}}$, the argument can be repeated,  over and over again.  The rates
of loss of $E$, $L_z$, and $Q$ will then continue at such a rate so as
to maintain circularity.

A more intuitive picture of why a  circular orbit remains circular was
provided by    Ori~\cite{ori},  who first     pointed  out  that   the
incommensurability of  the $r$ and $\theta$ periods  is the key reason
why  the  argument can  be made   without  knowing the nature  of  the
reaction  force:  Even  if  the radiation   reaction were  to take the
bizarre form of somebody with a hammer hitting the particle every time
the particle  is  at some value  of  $\theta$, there would have  to be
another person across  the  equatorial plane  at $\pi-\theta$  with  a
hammer hitting the particle in a corresponding way, as dictated by the
orbital symmetries.  Since  the $r-\bar r$  frequency  is not an  even
multiple  of   the    $\theta$ frequency,   the   hammer  hits  cannot
constructively interfere with each other and produce an eccentricity.

If an orbit is circular, then just knowing the  rates of change of $E$
and $L_z$   (for example, by knowing  the  energy and angular momentum
carried off in  the gravitational  waves)  is enough to determine  the
full orbital evolution since the evolution  of $Q$ is constrained such
that the conditions listed immediately before Eq.~(\ref{eq:rfunc}) are
satisfied, for as long as the orbit itself is stable.

\section{Leading order effect of spin on eccentric orbits}
\label{sec:reac}

We now wish to consider  general, not just  circular, orbits around  a
black  hole.     But in doing   so, we    restrict ourselves   to only
considering  the leading  order effect  of   spin.   We  will use  the
formalism of a    radiation reaction force   described in  a  previous
paper~\cite{circ} and merely state how the method as described in that
paper generalizes to orbits with eccentricity.

When one is only interested in  leading order in  $S$ and in $M/r$ (or
equivalently, $M/a$, in terms of orbit  parameters), the effect of the
hole's Kerr metric on the particle's motion can  be substituted with a
spin-orbit interaction in three dimensional flat-space.  Let spherical
polar coordinates  $r$,  $\theta$, and $\phi$,  centered  on the black
hole,  be used to    describe the  location   of the  particle  (these
coordinates describe the relative  separation of the two bodies), with
the  hole's    spin  along the  polar    axis.    The Lagrangian  {\bf
(}Ref.~\cite{kww}, Eq.~(4){\bf )}  for the motion  of  the particle is
given,   to  linear order in  $S$  but  otherwise  in solely Newtonian
theory, by
\begin{equation}
\label{eq:lagr}
{\cal L}  =
 \frac{\mu}{2} [\dot  r^2 + r^2 \dot  \theta^2 + r^2 \sin^2
	(\theta) \dot \phi^2]
 + \frac{\mu M}{r} - \frac  {2 \mu S \sin^2 \theta}{r} \dot \phi.
\end{equation}
To leading order
in $S$ and in $M/r$, the motion resulting from  this Lagrangian is the
same as in the Kerr metric.  The  use of flat space coordinates, which
ignores $M/r$ corrections, is  adequate to  leading order.  Using  the
same  coordinate variable  names $r$,  $\theta$, and  $\phi$ for these
coordinates as for the Kerr  metric's Boyer-Lindquist coordinates does
not cause conflict and should not  cause confusion.  Alternatively, we
can use Cartesian coordinates, $x_1= r\sin \theta
\cos \phi$, $x_2= r\sin \theta \sin \phi$, and $x_3 = r \cos \theta$.

The  Lagrangian~(\ref{eq:lagr})   admits three constants   of  motion,
called $E$, $L_z$, and $Q$ because  they are the  same constants as we
have in the Kerr  metric, to leading order in  $S$ and in $M/r$.   The
values of these constants are:
\begin{mathletters}
\label{eq:const}
\begin{eqnarray}
E=&&\frac{\mu}{2}[
\dot r^2 +r^2 \dot \theta^2 + r^2 \sin^2 (\theta) \dot
  \phi^2 ]-\frac{\mu M}{r},\\
L_z =&& \mu r^2 \sin^2(\theta)\dot \phi -\frac{2 \mu S \sin^2
\theta}{r},\\
Q+L_z^2=&&\mu^2r^4[\dot \theta^2 + \sin^2 (\theta) \dot \phi^2]
  - 4 \mu^2 S r \sin^2 (\theta) \dot \phi.
\end{eqnarray}
The combination $Q+L_z^2$ is a more natural constant to work with than
$Q$.  If $S$ were equal to zero, then $Q+L_z^2$ would be the square of
the total angular momentum.
\end{mathletters}

The constants of motion $\iota$, $e$, and $a$, when considered only to
leading order in $S$ and in $M/r$, are related to  $E$, $L_z$, and $Q$
by:
\begin{mathletters}
\label{eq:clo}
\begin{eqnarray}
\cos \iota =&& \frac{L_z}{\left(Q+L_z^2\right)^{1/2}},\\
1-e^2 =&& -2 \frac{E (Q+L_z^2)}{\mu^3 M^2}
	\left(1+4\frac{S M \mu^3 L_z}{(Q+L_z^2)^2}\right),\\
a =&&-\frac{M \mu}{2E} \left(1+2\frac{S M \mu^3
L_z}{(Q+L_z^2)^2}\right).
\end{eqnarray}
It is easy to verify these, by checking that the $\bar E$ and $\bar r$
that   would  make  Eqs.~(\ref{eq:defofiea}) give  Eqs.~(\ref{eq:clo})
satisfy (at leading  order in $S$  and in  $M/a$) the stable  circular
orbit constraints     listed immediately  before Eq.~(\ref{eq:rfunc}).
Note that  Eqs.~(\ref{eq:clo})  are valid for  arbitrary  eccentricity
$e$; they do not require $e \ll 1$.
\end{mathletters}

It  is possible to express the  instantaneous  time derivative of each
constant of motion, $dE/dt$,   $dL_z/dt$,  or $d(Q+L_z^2)/dt$,  as   a
function  of $r$, $x_3$,  $\dot r$,  $\dot x_3$, and  the constants of
motion; there is no occurrence  of $\phi$ (because of the axisymmetry)
or $\dot \phi$  (as this is determined  with $L_z$, $r$,  and $\theta$
known) in any of the expressions.   If $S$ were  zero then there could
be no $x_3$ dependence, rather only  $r$ dependence, since there is no
physically preferred direction when spin is absent.  Thus, an $x_3$ or
$\dot x_3$ can only show up  in a term that includes  a factor of $S$.
Because of this, to  compute the time  derivative of each constant  of
motion  to Newtonian order   plus the spin  correction,  $x_3(t)$ only
needs to be known to  Newtonian order, because  the spin correction to
$x_3$ would  be an $S^2$   term in the derivative  of  the constant of
motion.  On the  other hand, the radial motion  $r(t)$ of the particle
has to be   known to Newtonian order   plus the spin  correction.  The
$\phi(t)$ motion does not have  to be known at  all for computing  the
evolution of the constants of motion.

Let us, then,  compute $r$ and $x_3$  to the necessary orders.  One
of the Euler-Lagrange equations yields
\begin{equation}
\ddot r = - \frac{M}{r^2} + \frac{Q+L_z^2}{\mu^2 r^3} + 6\frac{S
L_z}{\mu r^4}.
\end{equation}
The solution of this, in terms of a parameter $\psi$, is
\begin{equation}
\label{eq:rtraj}
r=\frac{(Q+L_z^2)/(\mu^2M)}{1+e \cos \psi}
\left(1+\frac{S L_z \mu^3 M}{(Q+L_z^2)^2}(6+2e \cos \psi)\right),
\end{equation}
\begin{equation}
\frac{dt}{d\psi} = \frac{(Q+L_z^2)^{3/2}/(\mu^3 M^2)}{(1+e \cos
\psi)^2}
\left(1+6\frac{S L_z \mu^3 M}{(Q+L_z^2)^2}\right).
\end{equation}
In   the Newtonian limit of    $S=0$, these are    the equations for a
Keplerian ellipse, with the true anomaly~$\psi$ being the angle on the
orbital plane of the particle relative to periastron  as seen from the
hole.

To Newtonian order, $x_3 = r \cos \theta$ can be expressed as
\begin{equation}
x_3=r \sin \iota ~ \sin (\psi + \psi_0).
\end{equation}
Here, $\psi_0$ is some constant  that describes the orientation of the
ellipse on the orbital plane.  As seen  from the hole, $\psi_0$ is the
angle between the direction of  the periastron and the intersection of
the equatorial and orbital planes.

The orbital period, from periastron to periastron, is
\begin{equation}
\label{eq:period}
T=\int_0^{2 \pi} d\psi \frac{dt}{d\psi} =
 2 \pi M \left(\frac{\mu}{-2 E}\right)^{3/2}.
\end{equation}
It happens that  $T$,  when written in this   form, does not  have  an
explicit $S$ dependence.

This motion   we have  just   described is  that  in  the  absence  of
gravitational radiation reaction; now  we will  compute the effect  of
the radiation reaction acceleration.    We can take the equations  for
the rates of change  of $E$, $L_z$, and  $Q$ due to radiation reaction
for   a  particle going  around   a  more  massive  spinning body from
Eqs.~(10), (13), and (14)  of Ref.~\cite{circ}.  These  equations give
us   formulas  for $\dot   E$, $\dot  L_z$,   and  $d(Q +L_z^2)/dt$ as
functions of the displacement of the particle  relative to the hole in
Cartesian  coordinates, $x_k$, and the  relative  velocity $\dot x_k$.
There  will also  be higher order  time derivatives  of $x_k$ (such as
$\ddot x_k$, $\overdots  x_k$,  etc.), but  these  derivatives can  be
eliminated  from the expressions for  $\dot  E$, $\dot L_z$, and $\dot
{Q+L_z^2}$ with the   aid  of the  Euler-Lagrange   equations [derived
from~(\ref{eq:lagr})  when expressed   in Cartesian coordinates---note
that repeated indices are summed over 1,2,3]:
\begin{equation}
\ddot x_k = -\frac{M}{r^3} x_k +
	S \left(-\frac{4}{r^3}\epsilon_{3kj} \dot x_j +
	6\frac{\dot r}{r^4}\epsilon_{3kj}x_j + 6\frac{L_z}{\mu r^5}
x_k\right).
\end{equation}
The time evolution of each constant of motion can thereby be expressed
in terms  of $r$, $\dot r$,  $x_3$, $\dot x_3$,   and the constants of
motion.     The  trajectory~(\ref{eq:rtraj})--(\ref{eq:period}) can be
inserted into these expressions, and then time averaged using
\begin{equation}
\Bigl\langle \dot E \Bigr\rangle =\frac{1}{T} \int_{0}^{2 \pi}
d \psi \frac{dt}{d\psi} \dot E,
\end{equation}
and     similarly   for  $L_z$       and  $Q+L_z^2$.     The    result
is~\cite{mathematica}

\widetext
\begin{mathletters}
\label{eq:evolelq}
\begin{eqnarray}
\Bigl\langle \dot{E}\Bigr\rangle=&&
-\frac{32}{5} \frac{\mu^2}{M^2}
	\left(\frac{M}{a}\right)^5 \left(\frac{1}{1-e^2}\right)^{7/2}
 \biggl[\left(1 + \frac{73}{24} e^2 + \frac{37}{96}
e^4\right)\nonumber\\
&&~~~~~~-\frac{S}{M^2} \left(\frac{M}{a(1-e^2)}\right)^{3/2} \cos \iota
	\left(\frac{73}{12} + \frac{1211}{24} e^2 + \frac{3143}{96} e^4
	+ \frac{65}{64} e^6\right)\biggr],\\
\label{eq:evoll}
\Bigl\langle \dot{L_z}\Bigr\rangle=&&
-\frac{32}{5} \frac{\mu^2}{M} \left(\frac{M}{a}\right)^{7/2}
	\left(\frac{1}{1-e^2}\right)^2
	\Biggl[\cos \iota \left(1 + \frac{7}{8}e^2\right)
	+\frac{S}{M^2} \left(\frac{M}{a(1-e^2)}\right)^{3/2}\nonumber\\
&&~~~~~~\times \biggl(\left[\frac{61}{24} +
	 \frac{63}{8}  e^2 + \frac{95}{64} e^4\right] -\cos^2 \iota
	\left[\frac{61}{8} +\frac{109}{4}  e^2  +\frac{293}{64}
e^4\right]
	-\cos(2 \psi_0) \sin^2 \iota \left[\frac{5}{4} e^2 +
		\frac{13}{16} e^4\right]\biggr)\Biggr],\\
\Bigl\langle \dot{Q+L_z^2}\Bigr\rangle=&&
-\frac{64}{5}\mu^3 \left(\frac{M}{a}\right)^3
	\left(\frac{1}{1-e^2}\right)^{3/2}\biggl[\left(1 + \frac{7}{8}
    e^2\right)-\frac{S}{M^2}  \left(\frac{M}{a(1-e^2)}\right)^{3/2}\cos
\iota
	\left(\frac{97}{12} + 22 e^2 + \frac{99}{32} e^4\right)\biggr].
\end{eqnarray}
\end{mathletters}
The  evolution   of these    constants   can be    converted,    using
Eqs.~(\ref{eq:clo}), to the other set of constants:
\begin{mathletters}
\label{eq:evoliae}
\begin{eqnarray}
\label{eq:iotadot}
\Bigl\langle \dot{\iota}\Bigr\rangle=&&
\frac{\mu S}{M^4}
\left(\frac{M}{a}\right)^{11/2} \left(\frac{1}{1-e^2}\right)^4\sin
\iota
\left[\frac{244}{15} +\frac{252}{5}e^2+\frac{19}{2}e^4-\cos(2\psi_0)
\left(8e^2+\frac{26}{5}e^4\right)\right],\\
\label{eq:adot}
\Bigl\langle \dot{a}\Bigr\rangle=&&
-\frac{64}{5} \frac{\mu}{M} \left(\frac{M}{a}\right)^3
\left(\frac{1}{1-e^2}\right)^{7/2}
\biggl[\left(1 + \frac{73}{24} e^2 + \frac{37}{96} e^4\right)
\nonumber\\
&&~~~~~~- \frac{S}{M^2} \left(\frac{M}{a(1-e^2)}\right)^{3/2}\cos \iota
\left(\frac{133}{12}+\frac{337}{6}e^2+\frac{2965}{96}e^4+
	\frac{65}{64}e^6\right)\biggr],\\
\label{eq:edot}
\Bigl\langle \dot{e}\Bigr\rangle=&&
-\frac{\mu}{M^2} \left(\frac{M}{a}\right)^4
\left(\frac{1}{1-e^2}\right)^{5/2} e
\biggl[\frac{304+121e^2}{15} - \frac{S}{M^2}
\left(\frac{M}{a(1-e^2)}\right)^{3/2}\cos \iota~
\left(\frac{1364}{5}+\frac{5032}{15}e^2+\frac{263}{10}e^4\right)\biggr]
{}.
\end{eqnarray}
\end{mathletters}
\narrowtext

Equations~(\ref{eq:evolelq})  agree (after trivial conversions of
notation)
with  previous results: Eqs.~(15)  of  Ref.~\cite{circ} and (the first
line of) Eq.~(3.14) of Ref.~\cite{shibata}, each of which is a special
case of Eqs.~(\ref{eq:evolelq}).

In  most cases,  the terms with   the $\cos(2\psi_0)$  can  be dropped
because they average to zero; to  see when this  can be done, consider
the following: The Newtonian approximation  to the motion is that  the
particle travels in  an ellipse.  The  first correction to this motion
is, as Einstein computed for  Mercury, that the periastron position of
the ellipse shifts on a timescale of
\begin{equation}
T_{\rm prec} \sim M(M/a)^{-5/2}(1-e^2).
\end{equation}
The radiation reaction timescale for  terms that involve $\psi_0$,  as
computed     by        evaluating         $(\sin    \iota)/\langle\dot
\iota\rangle_{\psi_0~{\rm terms}}$, is
\begin{equation}
T_{\rm     rad} \sim   M
\left(\frac{M}{a}\right)^{-11/2}\left(\frac{S}{M^2}\right)^{-1}\left(
\frac{\mu}{M}\right)^{-1}\frac{(1-e^2)^4}{e^2}.
\end{equation}
(There  are also factors of  order unity that  involve  $e$ which were
ignored.  If  $T_{\rm rad}$ were  computed differently, for example by
evaluating  $L_z/\langle\dot L_z   \rangle_{\psi_0~{\rm terms}}$,   it
would contain factors of $\iota$ as well.)

In  the Newtonian limit,  $\psi_0$ is  fixed,  but with the periastron
precession,  $\psi_0$   changes  slightly   after  each orbit,   by  a
post-Newtonian correction  that was ignorable  until now: When $T_{\rm
rad} \gg T_{\rm prec}$,  the $\cos(2\psi_0)$  in Eqs.~(\ref{eq:evoll})
and    (\ref{eq:iotadot}) averages to zero,  and   the terms with that
factor can be dropped.   For extremely eccentric orbits, $T_{\rm rad}$
might not be  much greater than $T_{\rm prec}$,  so the $\psi_0$ terms
must be kept.  In all other respects, the periastron precession can be
ignored because  it just gives  terms higher order  in $M/a$ (terms we
have  neglected).   The only  effect of  the precession,  to which our
analysis is sensitive, is  the averaging away  of $\psi_0$ in the case
that $T_{\rm rad} \gg T_{\rm prec}$.

{}From  Eq.~(\ref{eq:iotadot}), it is   clear  that the angle  $\iota$
changes such   as   to   become  antialigned   with    the  spin.   In
Ref.~\cite{circ},  this conclusion   was reached for  circular orbits;
finite  eccentricity does not  change, but only enhances, this result.
However,  the statement that  ``the  inclination angle antialigns with
the spin'' is  subject to  the  warning that we  mentioned above  when
introducing $\iota$: With the orbit not confined to a fixed plane, the
angle $\iota$  is  not the  only   way we could  define  ``inclination
angle''~\cite{circ}.

Equation~(\ref{eq:edot})  has  two  important consequences:  First, to
leading order, orbits  tend to circularize, as  is  a well-known fact.
Second, if an orbit is circular, then $e=0$ and $\langle\dot e \rangle
= 0$, so the orbit remains circular.  This is  expected, since this is
the leading order limit of the general result in Sec.~\ref{sec:orb}.

The above  analysis    is just one  step   in  a general  program  for
understanding the effects  of radiation reaction on orbiting, spinning
bodies.    Future steps  in  this program    include: generalizing the
analysis to an arbitrary  mass ratio $\mu/M$  and to the case of  both
masses having  spin, extending the analysis  to higher  order in $M/r$
and   in $S$,  and  achieving a  similar calculation   of  the orbital
evolution in the fully relativistic Kerr metric.

 \acknowledgements{The  author is grateful  to  Daniel Kennefick, Alan
Wiseman, and Kip Thorne for their  advice.  This work was supported by
NSF Grant AST-9417371, and by NASA Grant NAGW-4268.}

\end{document}